\pdfoutput=1
\RequirePackage{ifpdf}
\ifpdf 
\documentclass[pdftex]{sigma}
\else
\documentclass{sigma}
\fi

\numberwithin{equation}{section}

\newcommand{\be}{\begin{equation}}
\newcommand{\ee}{\end{equation}}

\begin{document}
\allowdisplaybreaks

\renewcommand{\thefootnote}{}

\newcommand{\arXivNumber}{2304.07895}

\renewcommand{\PaperNumber}{054}

\FirstPageHeading

\ShortArticleName{Moduli Space for Kink Collisions with Moving Center of Mass}

\ArticleName{Moduli Space for Kink Collisions\\ with Moving Center of Mass\footnote{This paper is a~contribution to the Special Issue on Topological Solitons as Particles. The~full collection is available at \href{https://www.emis.de/journals/SIGMA/topological-solitons.html}{https://www.emis.de/journals/SIGMA/topological-solitons.html}}}

\Author{Christoph ADAM~$^{\rm a}$, Chris HALCROW~$^{\rm b}$, Katarzyna OLES~$^{\rm c}$, Tomasz ROMANCZUKIEWICZ~$^{\rm c}$ \newline and Andrzej WERESZCZYNSKI~$^{\rm c}$}

\AuthorNameForHeading{C.~Adam, C.~Halcrow, K.~Oles, T.~Romanczukiewicz and A.~Wereszczynski}

\Address{$^{\rm a)}$~Departamento de F\'isica de Part\'iculas, Universidad de
Santiago de Compostela and \\
\hphantom{$^{\rm a)}$}~Instituto Galego de F\'isica de Altas Enerxias (IGFAE), E-15782
Santiago de Compostela,\\
\hphantom{$^{\rm a)}$}~Spain}
\EmailD{\href{mailto:adam@igfae.usc.es}{adam@igfae.usc.es}}

\Address{$^{\rm b)}$~Department of Physics, KTH-Royal Institute of Technology, SE-10691 Stockholm, Sweden}
\EmailD{\href{mailto:chalcrow@kth.se}{chalcrow@kth.se}}

\Address{$^{\rm c)}$~Institute of Theoretical Physics, Jagiellonian University,
Lojasiewicza 11, Krak\'{o}w, Poland}
\EmailD{\href{mailto:katarzyna.slawinska@uj.edu.pl}{katarzyna.slawinska@uj.edu.pl}, \href{mailto:tomasz.romanczukiewicz@uj.edu.pl}{tomasz.romanczukiewicz@uj.edu.pl},\newline \hspace*{18mm}\href{mailto:andrzej.wereszczynski@uj.edu.pl}{andrzej.wereszczynski@uj.edu.pl}}

\ArticleDates{Received April 20, 2023, in final form July 26, 2023; Published online August 02, 2023}

\Abstract{We apply the collective coordinate model framework to describe collisions of a~kink and an antikink with nonzero total momentum, i.e., when the solitons possess different velocities. The minimal moduli space with only two coordinates (the mutual distance and the position of the center of mass) is of a wormhole type, whose throat shrinks to a point for symmetric kinks. In this case, a singularity is formed. For non-zero momentum, it prohibits solutions where the solitons pass through each other. We show that this unphysical feature can be cured by enlarging the dimension of the moduli space, e.g., by the inclusion of internal modes.}


\Classification{35C08; 35Q51}

\renewcommand{\thefootnote}{\arabic{footnote}}
\setcounter{footnote}{0}

\section{Motivation}
Recently, there has been significant progress in the construction of collective coordinate models (CCM) for kink collisions in (1+1) dimensions.
After more than forty years of struggle~\cite{CSW, Kev, sug, Weigel}, a CCM qualitatively describing symmetric kink-antikink scattering in $\phi^4$ theory was presented in~\cite{MORW}. As expected for a long time, it relies on the resonant energy transfer mechanism~\cite{CSW}, which allows for an energy flow between the kinetic and internal degrees of freedom (DoF) which, in this case, are reduced to a single normal mode, the so called shape mode. In this CCM approach it was crucial to take into account two new ingredients. Namely, a suitable regularization of the null vector problem~\cite{MORW-2} and a proper specification of the initial conditions for the CCM~\cite{MORW}. Importantly, the last issue ensures an approximate modeling of the boosted kink.

One reason to construct accurate CCMs is that they allow for a semiclassical quantisation of solitons, which, despite the recent progress of the more direct approach~\cite{JE-1}, is still the simplest method of quantisation. If the approximation captures the nonlinear dynamics classically, these effects will also be captured in the quantisation. Soliton quantisation has been carried out for Bogomolny--Prasad--Sommerfield (BPS) models such as critically coupled vortices~\cite{BPS-moduli-2} and monopoles~\cite{BS-mono}. But it is much more difficult for non-BPS systems, which contain forces between solitons. The prototypical example is the quantisation of skyrmions as nuclei, with only a few examples which go beyond a rigid rotor with harmonic vibrations approximation~\cite{CH-B7, RL-deut}. The key difficulty is the construction of good collective coordinates, rather than the quantisation itself. Indeed, it was recently shown that the best CCM for skyrmions, the instanton approximation, does not even describe asymptotic skyrmion dynamics correctly~\cite{CHDH}. Hence, there is need for a better understanding of CCMs for non-BPS systems. The kink-antikink models discussed in this paper are in that category. Moreover, an accurate approximation for soliton-antisoliton dynamics can be used to improve the calculation of, e.g., monopole-antimonopole pair production~\cite{IA-mono}.

The previous paragraph provides motivation to carefully study CCM models. The CCM for~$\phi^4$ theory has recently been extended to the perturbative relativistic CCM (pRCCM), which is a tool allowing for a precise quantitative description of (at least some) kink collisions within the CCM framework~\cite{RMS}. In this extension, there are an arbitrary number of internal degrees of freedom known as Derrick modes. Further generalizations are sometimes needed if, during a collision, so-called delocalized, i.e., two-soliton normal modes, are formed. This may happen in collisions of kinks interpolating between vacua with different masses of small perturbations (mesons) as, e.g., in kink-antikink collisions in the $\phi^6$ theory~\cite{Tr}.

In all previously considered kink-antikink processes the solitons are boosted towards each other with {\it the same velocity}, that is, their center of mass (CoM) is stationary. This is also true for more general scatterings where internal modes have also been excited~\cite{AI}. This is an obvious simplification since motion of the center of mass is trivial for relativistic field theory. However, we will show that at the level of a CCM it is not trivial as it can add a singularity to our space. In the current paper, we will investigate this issue and resolve it by adding a new collective coordinate to our model.

Our first result is that the moduli space of minimal dimension, $\mathcal{M}(\alpha,\beta)$, where we include only the intersoliton distance $\alpha$ and the position of the center of mass $\beta$, has the shape of a~wormhole. Further, the minimum radius of the throat reflects how close the set of configurations spanning the moduli space gets to the vacuum solution. In the symmetric case, where the antikink is simply minus the kink, the vacuum belongs to the moduli space. Therefore, the throat shrinks to a point and geometrically the moduli space takes the form of two trumpets (for $\alpha>0$ and $\alpha<0$, respectively) which are glued together at $\alpha=0$. In other words, a~singularity emerges. For nonzero momentum, i.e., where the initial kink and antikink have different velocities, this singularity prohibits solutions to move from one trumpet-like part to the other. So, for nonzero momentum, the resulting CCM does not allow the solitons to pass through each other, even when their velocities are infinitesimally close, that is, when the center of mass moves arbitrarily slowly. This is an unphysical feature, not observed in collisions in the full field theory. \looseness=1

We explain these results using an analogy to point particle motion on a two-dimensional plane in the presence of a radially symmetric potential.

Secondly, we show how this unphysical feature is remedied if we extend the moduli space by the inclusion of internal modes, such as the shape mode in the case of $\phi^4$ kinks. The singularity remains but, after resolving the usual null-vector problem, its dimensionality drops from three to two in a four-dimensional moduli space. Hence there exist solutions of the CCM where $\alpha$ can change sign, describing solitons that pass through each other. In the extended moduli space, this is made possible by the excitation of the shape modes carried by the solitons. \looseness=1 \newpage

\section{Singularity of the minimal moduli space}\label{sec2}
Let us consider a generic scalar field theory in (1+1) dimensions
\begin{gather}
\mathcal{L}=\frac{1}{2}\phi_t^2-\frac{1}{2}\phi_x^2-U(\phi), \label{Lag}
\end{gather}
where we assume that the potential $U$ has at least two minima, where it takes the value 0. The kink $\Phi_K(x-a)$ and antikink $\Phi_{\bar{K}}(x-a)$ are solutions of the pertinent first order Bogomolny equation
\[
\frac{{\rm d} \phi}{{\rm d}x} = \pm \sqrt{2U(\phi)}
\]
and saturate the topological energy bound in each topological sector. Here, $a$ denotes the position of the soliton which can be identified with, e.g., the zero of the field for $\phi^4$ theory.

The CCM framework allows for the analysis of solitonic processes in a given field theory, that is, a system with infinitely many DoF, in terms of a finite-dimensional dynamical system, see~\cite{MS} for details. This is obtained by the identification of a restricted set of configurations $\mathcal{M}=\big\{\Phi\bigl(x; X^i\bigr), i=1,\dots,N \big\}$ which are meant to cover the main features of the process in question. Such static configurations are parametrized by a {\it finite} set of parameters, {\it moduli}, $X^i$ which are promoted to time dependent collective coordinates. Inserting these configurations into the original Lagrangian (\ref{Lag}) and performing spatial integration results in a corresponding CCM governed by the Lagrangian
\begin{gather*}
L[{\bf X}]=\int_{-\infty}^\infty \mathcal{L}\big[\Phi\bigl(x; X^i(t)\bigr)\big] \, {\rm d}x
= \frac{1}{2} g_{ij}({\bf X}) \dot{X}^i \dot{X}^j - V({\bf X})  ,
\end{gather*}
where
\begin{gather*}
g_{ij}({\bf X})=\int_{-\infty}^\infty \frac{\partial \Phi}
{\partial X^i} \frac{\partial \Phi}{\partial X^j} \, {\rm d}x
\end{gather*}
is the metric on $\mathcal{M}$, while
\begin{gather*}
V({\bf X})=\int_{-\infty}^\infty \bigg( \frac{1}{2}
\bigg( \frac{\partial \Phi}{\partial x}
\bigg)^2 + U(\Phi) \bigg) \, {\rm d}x
\end{gather*}
is the potential energy.

The CCM approach works well not only for BPS models~\cite{Bogomolny}, where there is no static force between solitons~\cite{BPS-moduli-1,NM-2, NM-1, BPS-moduli-2}, but also in strongly non-BPS systems like kink-antikink scattering in $\phi^4$~\cite{MORW} and $\phi^6$~\cite{phi6} theories. It provides a tool for a semi-analytical analysis of soliton interactions explaining, e.g., the $\pi/2$ scattering typical for many soliton processes. It also provides a strong argument, if not a proof, that the chaotic structure in kink collisions in the $\phi^4$~\cite{RMS, MORW} and $\phi^6$~\cite{phi6} models is due to the resonant energy transfer mechanism. As mentioned in the Introduction, all CCM considered so far assumed that both solitons collide with equal velocities and, therefore, the CoM does not move.

Here we relax this assumption and will consider kink-antikink collisions where the solitons have {\it different} initial velocities, which induce nontrivial motion of the center of mass. On the level of the CCM, this can be achieved by assuming that each soliton has an independent position. Therefore, for the study of kink-antikink collisions we propose the following set of restricted configurations
\[
\Phi_{K\bar{K}}(x;a,b)=\Phi_K(x-a)+\Phi_{\bar{K}}(x-b) +\Phi_v,
\]
where $\Phi_v$ is a pertinent constant (vacuum value), which provides the proper boundary conditions. This is obviously the simplest choice ignoring, e.g., internal modes, which are crucial for the correct description of scattering processes. However, in the zero-momentum case ($b=-a$), it is a~well understood starting point. We remark that instead of the exact kink and antikink fields one can use some approximations based on, e.g., $CP^1$ instantons~\cite{cp-inst} or, better, constrained instantons~\cite{c-inst}. It is convenient to immediately introduce another set of collective coordinates~$\alpha$ and~$\beta$ which describe, respectively, half the distance between the solitons and the CoM: ${a=-\alpha+\beta}$, $b=\alpha +\beta$. Thus,
\[
\Phi_{K\bar{K}}(x;\alpha, \beta)=\Phi_K(x+\alpha-\beta)+\Phi_{\bar{K}}(x-\alpha-\beta) +\Phi_v.
\]
The resulting two-dimensional moduli space $\mathcal{M}(\alpha,\beta)$ has the following metric functions
\begin{gather*}
g_{\alpha\alpha}=\int_{-\infty}^\infty \bigl( \Phi_K'(x+\alpha-\beta) - \Phi_{\bar{K}}'(x-\alpha-\beta) \bigr)^2 \,{\rm d}x= 2M+2\tilde{M}(\alpha),
\\
g_{\beta \beta}=\int_{-\infty}^\infty \bigl( \Phi_K'(x+\alpha-\beta) + \Phi_{\bar{K}}'(x-\alpha-\beta) \bigr)^2 \,{\rm d}x= 2M-2\tilde{M}(\alpha) .
\end{gather*}
Here the prime denotes derivative with respect to the argument. In addition, $M$ is the mass of the solitons
\[
M=\int_{-\infty}^\infty \Phi_K^{'2} (x) {\rm d}x =\int_{-\infty}^\infty \Phi_{\bar{K}}^{'2} (x)\, {\rm d}x
\]
and $\tilde{M}(\alpha)$ is a nontrivial contribution due to the intersoliton interaction
\[
\tilde{M}(\alpha)=-\int_{-\infty}^\infty \Phi_K'(x-\alpha) \Phi_{\bar{K}}'(x+\alpha)\, {\rm d}x.
\]
The off-diagonal component of the metric $g_{\alpha\beta}$ vanishes identically. Due to translational symmetry, the metric does not depend on the position of the center of mass $\beta$.

The kink (antikink) is always given by a monotonically growing (decreasing) function with $\Phi'_K(x)>0$ and $\Phi'_{\bar{K}}(x) <0$. This is true for all single component kink theories. Hence we can conclude that $\tilde{M}(\alpha)>0$. Furthermore, $\tilde{M}(\alpha) \leq M$, due to the non-negativity of the metric.

Now, we assume that there exists an $\alpha_0$ for which $\tilde{M}$ takes a maximal value, which is {\it smaller than the mass of the kink}, $\tilde{M}_0=\tilde{M}(\alpha_0)<M$. Thus in the vicinity of this point
\[
\tilde{M}(\alpha)= \tilde{M}_0 - \tilde{M}_2 (\alpha_0-\alpha)^2 +o\bigl((\alpha-\alpha_0)^2\bigr)
\]
and the corresponding metric reads
\[
{\rm d}s^2 = 2\bigl(M+\tilde{M}_0\bigr){\rm d}\alpha^2 + 2 \bigg(M-\tilde{M}_0 + \frac{1}{2}\tilde{M}_2 (\alpha_0-\alpha)^2 \bigg) {\rm d}\beta^2.
\]
This is a {\it wormhole} type metric, with the throat at $\alpha=\alpha_0$ and with its throat radius squared $R^2=2\bigl(M-\tilde{M}_0\bigr)$.

Let us now focus on the extremal case where $\tilde{M}(\alpha_0)=M$, when the metric function $g_{\beta \beta}$ vanishes. This requires
\[
-\int_{-\infty}^\infty \Phi_K'(x-\alpha_0) \Phi_{\bar{K}}'(x+\alpha_0) \,{\rm d}x= \int_{-\infty}^\infty \Phi_K'(x) \Phi_{K}'(x) \,{\rm d}x,
\]
which happens for $\alpha_0=0$, as the kink and antikink are related by a simple sign multiplication
\be
\Phi_{\bar{K}}(x)=-\Phi_K(x). \label{symK} \,
\ee
We call these {\it symmetric solitons}. They occur if, e.g., the field theoretical potential has {\it only} two vacua and enjoys the reflection symmetry $U(-\phi)=U(\phi)$ which occurs in $\phi^4$ theory. A different example of such kinks is provided by the sine-Gordon model.

\begin{figure}[t]\centering
\includegraphics[width=0.8\columnwidth]{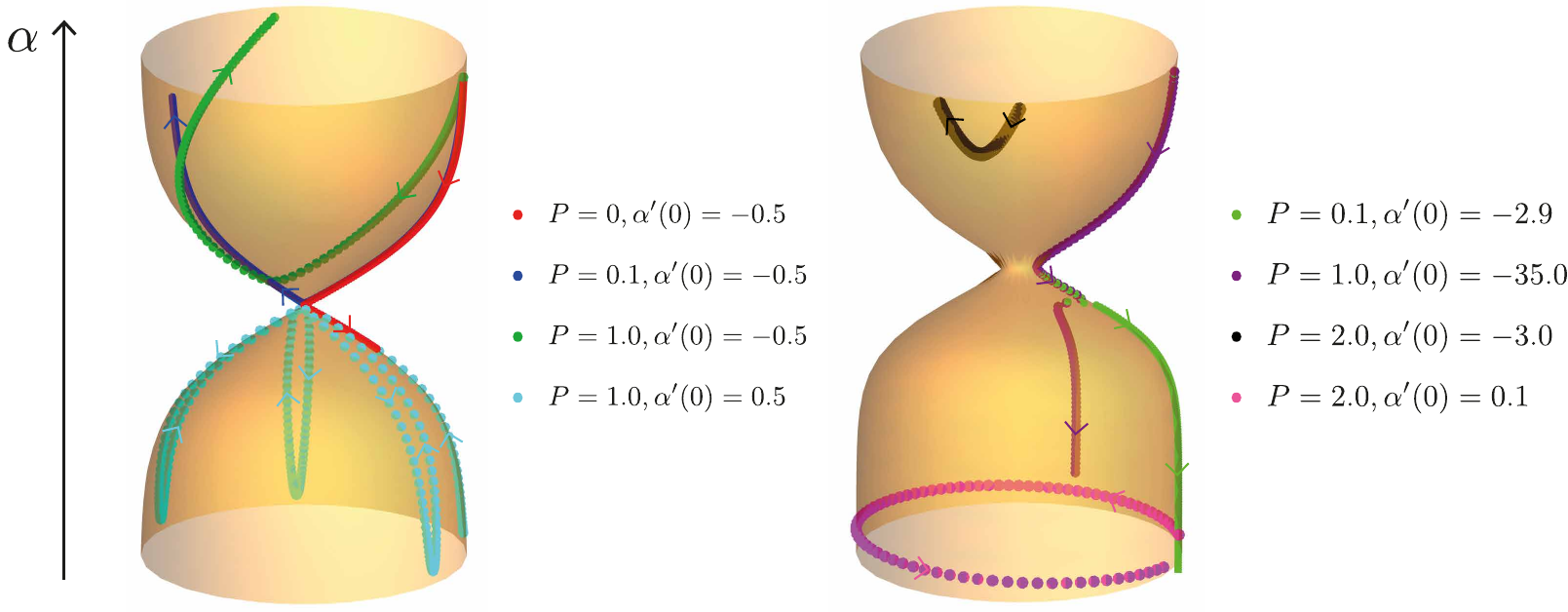}
 \caption{The 2-dimensional collective coordinate space for the $\phi^4$ theory (left) and $\phi^6$ theory (right), isometrically embedded in $\mathbb{R}^3$ as a surface of revolution. The spaces have singular and wormhole metrics. Four numerically generated trajectories are plotted in red, blue, dark green and teal with initial data $(\alpha, \dot{\alpha}, P)$ equal to $(3, -0.5, 0)$, $(3, -0.5, 0.1)$, $(3, -0.5, 1.0)$ and $(-3, 0.5, 1.0)$ and in light green, purple, black and pink with initial data $(\alpha, \dot{\alpha}, P)$ equal to $(3, -2.9, 0.1)$, $(3, -35.0, 1.0)$, $(3, -3.0, 2.0)$ and $(-3, 0.1, 2.0)$. Observe that on the left the trajectories with non-zero $P$ cannot pass through the origin, and remain in their initial half of moduli space. Only the zero-momentum path (red) is able to pass through the point $\alpha = 0$. On the right, trajectories can pass through the wormhole throat.} \label{MS-plot}
 \end{figure}

We can embed the collective coordinate model in $\mathbb{R}^3$ to visualise it, and do so for CCMs of the $\phi^4$ and $\phi^6$ models in Figure \ref{MS-plot}. We also plot numerically generated scattering trajectories on the spaces. The most interesting point is $\alpha=0$, and we can write the metric near here explicitly
\[
{\rm d}s^2_{\rm sym} = 4M{\rm d}\alpha^2 + \tilde{M}_2 \alpha^2{\rm d}\beta^2 .
\]
This is the metric of the plane with origin removed where $\beta$ plays the role of the polar angle, but with infinite range. Near $\alpha=0$ the space is an infinite-sheeted version of a flat double cone and is singular~\cite{RMS}. We can explicitly see the singularity in the $\phi^4$ model at $\alpha=0$. This point splits the space in half and trajectories cannot pass through it if they have non-zero $P$. In contrast, the $\phi^6$ theory has no singularity but is instead a smooth wormhole with a curved throat.

It is instructive to see that for symmetric solitons the point $\alpha=0$ corresponds to the {\it exact vacuum} configuration. Indeed \eqref{symK} implies that $\Phi^{\rm sym}_{K\bar{K}}(x;\alpha=0,\beta)=\Phi_v$. For non-symmetric solitons
\[
\Phi_{K\bar{K}}(x;\alpha=0,\beta)=\Phi_K(x-\beta)+\Phi_{\bar{K}}(x-\beta) +\Phi_v,
\]
which is non-zero. The collapse of the wormhole and appearance of the singularity for the symmetric solitons is related to the fact that the restricted set of configurations contains the exact vacuum solution. The metric component $g_{\beta\beta}$ is zero because at the vacuum $\alpha=0$, configurations with different $\beta$ are equal and hence there is zero distance between them. For non-symmetric kinks all configurations on the moduli space $\mathcal{M}$ are different from the vacuum and from each other, which keeps the wormhole throat open. The radius of the throat measures how close the closest configuration gets to the vacuum.

It is quite intriguing to observe that the singularity we found is {\it identical} to the previously identified singularity of the two-dimensional {\it relativistic moduli space} for kink-antikink collisions with zero total momentum. Again, this singularity is formed for symmetric kinks. To show this, let us consider the pertinent restricted set of configurations
\[
\Phi_{K\bar{K}}(x; \alpha, \lambda) = \Phi_K(\lambda(x+\alpha))-\Phi_K(\lambda(x-\alpha)) +\Phi_v,
\]
where, in addition to the position of the solitons, $\alpha$, we include the scaling (Derrick) deformation~$\lambda$. In the single soliton sector, the inclusion of the Derrick deformation reproduces the Lorentz contraction of the moving (anti)kink and, therefore, elevates the CCM from a non-relativistic to a relativistic point particle system~\cite{Rise}. To analyze the issue of the singularity, it is convenient to introduce a new coordinate $\tilde{\alpha}=\lambda \alpha$. Thus
\[
\Phi_{K\bar{K}}(x; \tilde{\alpha}, \lambda) = \Phi_K(\lambda x+\tilde{\alpha})-\Phi_K(\lambda x-\tilde{\alpha}) +\Phi_v,
\]
which for $\tilde{\alpha}\to 0$ at linear order gives
\[
\Phi_{K\bar{K}}(x; \tilde{\alpha}, \lambda) = \Phi_v+ 2\tilde{\alpha} \Phi'_K(\lambda x).
\]
This results in the following metric on the moduli space
\[
{\rm d}s^2= 4M\bigg( S \frac{\tilde{\alpha}^2}{\lambda^3}{\rm d}\lambda^2 - \frac{\tilde{\alpha}}{\lambda^2}{\rm d}\lambda {\rm d}\tilde{\alpha} + \frac{1}{\lambda} \tilde{\alpha}^2 \bigg),
\]
where
\[
S= \int_{-\infty}^\infty x^2 \Phi_K'^2(x)\,{\rm d}x \bigg/ \int_{-\infty}^\infty \Phi_K'^2(x)\,{\rm d}x.
\]
This non-diagonal metric admits a diagonal form after another change of variables, $\sigma = \tilde{\alpha} / \sqrt{\lambda}$ and $\tau=\ln (\lambda)$,
\[
{\rm d}s^2= 4M \bigg( {\rm d}\sigma^2 +\bigg(S-\frac{1}{4} \bigg) \sigma^2 {\rm d}\tau^2 \bigg).
\]
This is nothing but a version of a metric on the plane with the origin removed and with the coordinate $\tau$ playing the role of an infinitely extended angle. Hence, we get again a double cone glued at $\tilde{\alpha}=0$. Note that the singularity once again arises due to the fact that we pass through the exact vacuum, where the distance between configurations with different $\lambda$ is zero.
\section{Conserved momentum and reduced moduli spaces}\label{sec3}
In this section, we focus on two-dimensional CCMs with symmetric solitons and, hence, a~degenerate metric. We will show that the two parts of the moduli space ($\alpha>0$ and $\alpha<0$) are dynamically separated provided there is CoM motion. To show it we explicitly write the obtained CCM
\[
L(\alpha,\beta) = \frac{1}{2} g_{\alpha\alpha}(\alpha) \dot{\alpha}^2 + \frac{1}{2} g_{\beta \beta}(\alpha) \dot{\beta}^2 - V(\alpha).
\]
The effective potential $V(\alpha)$ is a smooth function of the intersoliton distance only, $V=V(\alpha)$. Its actual form depends on the field theoretical potential $U(\phi)$. Asymptotically, for ${\alpha \to \infty}$ where the kink and antikink are infinitely separated, it tends to twice the soliton mass, ${V(\alpha =\infty)\!=\!2M}$, and is zero for the vacuum configuration, $V(\alpha=0)=0$. For negative $\alpha$, which describes the regime where the solitons have passed through each other, the effective potential can grow to arbitrarily large values as $\alpha \to -\infty$, as happens for $\phi^4$ theory. If the field theory has more than two vacua, the effective potential $V$ may tend to a constant value, being again twice the mass of the pertinent solitons. This occurs, e.g., in the sine-Gordon model.

It is straightforward to observe that this CCM gives rise to two conserved quantities: the energy
\[
E=\frac{1}{2} g_{\alpha\alpha}(\alpha) \dot{\alpha}^2 + \frac{1}{2} g_{\beta \beta}(\alpha) \dot{\beta}^2 +V(\alpha)
\]
and the momentum
\[
P=\frac{{\rm d} L}{{\rm d} \dot{\beta}} = g_{\beta \beta}(\alpha) \dot{\beta}.
\]
The existence of the conserved momentum is related to the invariance of the CCM under a shift of $\beta$, which is ultimately due to translational invariance.

We can use the conserved momentum to eliminate the $\beta$ coordinate completely from the CCM. This brings us to a family of reduced CCMs defined on a one-dimensional moduli space~$\mathcal{M}(\alpha)$ equipped with the reduced effective potential $V_{\rm red}(P,\alpha)$ labeled by the momentum $P$
\[
L(\alpha) = \frac{1}{2} g_{\alpha\alpha}(\alpha) \dot{\alpha}^2 - V_{\rm red}(P, \alpha),
\]
where
\[
V_{\rm red}(P, \alpha)= V(\alpha)+\frac{P^2}{2 g_{\beta \beta} } .
\]
When $P=0$, the one-dimensional (reduced) moduli space is free of singularities. However, when $P\neq 0$, the reduced effective potential $V_{\rm red}$ reveals a singularity at $\alpha=0$. At this point, the metric function $g_{\beta\beta}$ is zero. Thus, such a smooth moduli space is dynamically divided into {\it two disconnected parts}, owing to the existence of the infinite potential barrier at $\alpha=0$. Note that there is no smooth transition to the zero momentum limit. There, the singularity of the reduced effective potential disappears and the whole moduli space is allowed. This demonstrates that the two parts of the moduli space with $\alpha>0$ and $\alpha<0$ are dynamically separated if the CoM has nonzero momentum.

It is instructive to compare the CCM for the symmetric solitons with a massive point particle in the $(r, \phi)$ plane in the presence of a central potential
\[
L_{\rm point} (r,\phi) = \frac{m}{2} \bigl( \dot{r}^2+r^2\dot{\theta}^2 \bigr) - V(r).
\]
There are two conserved quantities, the energy
\[
E_{\rm point}= \frac{m}{2} \bigl( \dot{r}^2+r^2\dot{\theta}^2 \bigr) - V(r)
\]
and the angular momentum
\[
S_{\rm point}=m\dot{r}.
\]
For nonzero angular momentum, $S_{\rm point}>0$, the point particle cannot reach the origin. This is not a problem for $r \in \mathbb{R}_+$, but it is for the colliding kinks. In this case, the distance $\alpha$ between the kink and antikink plays the role of the radial coordinate $r$ while the CoM $\beta$ can be viewed as an infinite-sheeted version of the angular variable $\theta$. The main difference is that $\alpha \in \mathbb{R}$ and therefore, there are two, positive and negative, parts. Furthermore, the momentum of the CoM plays the role of the angular momentum of the point particle. In any case, as the ``origin" $\alpha=0$ is forbidden due to the conservation laws for $P\neq 0$, these two parts separate whenever the momentum of the CoM is not zero.

A similar, but less extreme, problem occurs for CCMs which almost contain the vacuum, such as the two-dimensional $\phi^6$ space (seen in the right of Figure~\ref{MS-plot}). Here, the effective potential also contains $P^2/2g_{\beta \beta}$ which, in this case, is large since $g_{\beta\beta}$ is small. So, the space is not disconnected, but the wormhole throat provides a large effective potential barrier between its two halves when $P\neq0$. This is why the initial velocities of the paths which cross the wormhole are so large in Figure~\ref{MS-plot}. In reality, we do not expect such large velocities to be needed and hence this model contains similar problems to the singular, symmetric case. Hence the singularity is a worst-case symptom of a problem which afflicts a wide set of CCMs: any models which \emph{almost} contains the vacuum.

We remark that the appearance of the singularity for the case of symmetric kinks does not seem to be related to the fact that the CCM does not preserve the relativistic invariance of the original theory. First of all, the singularity even exists for arbitrarily small nonzero values of the momentum $P$ and, therefore, concerns the non-relativistic regime. In addition, as we have shown above, a similar singularity is observed also in a CCM which reproduces the relativistic contraction of the solitons. It is rather the passing through the exact vacuum which produces the singularity.
\section{Examples}
\subsection[KAK collisions in the phi\^{}4 and sine-Gordon models]{KAK collisions in the $\boldsymbol{\phi^4}$ and sine-Gordon models}
\begin{figure}[t]\centering
\includegraphics[width=0.45\columnwidth]{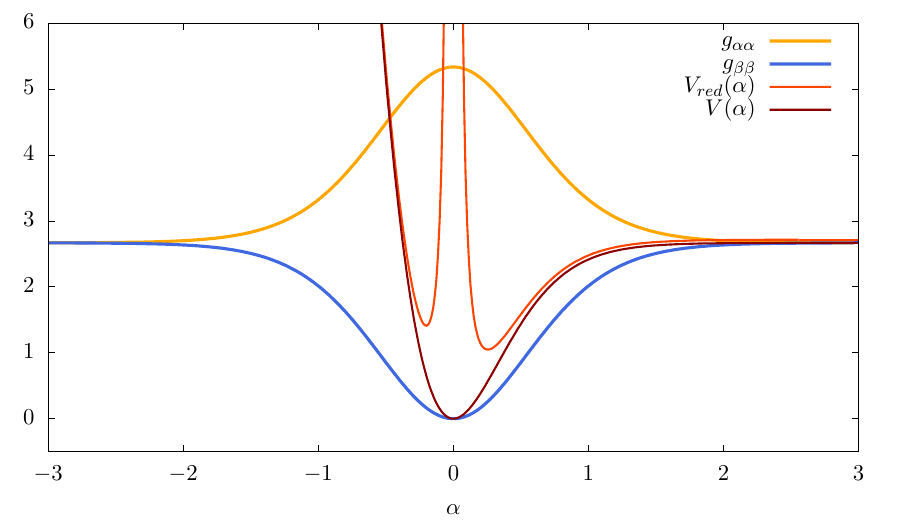}\qquad
\includegraphics[width=0.45\columnwidth]{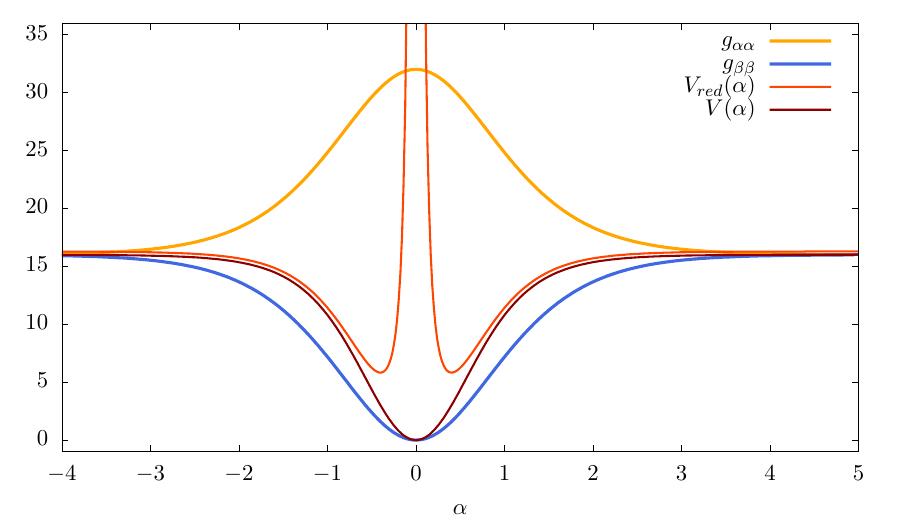}
 \caption{The metric functions $g_{\alpha\alpha}$ (orange), $g_{\beta\beta}$ (blue), the effective potential $V(\alpha)$ (brown) and the reduced effective potential $V_{\rm red}$ (red) for kink-antikink collisions. Left: $\phi^4$ model $(P=0.5)$; Right: sine-Gordon model ($P=3$).} \label{metric-plot}
 \end{figure}

In this subsection, we will study the above introduced construction for the case of symmetric kinks. We begin with the $\phi^4$ model defined by the field theoretical potential
\[
U=\frac{1}{2} \bigl(1-\phi^2\bigr)^2,
\]
which has the well-known kink solution $\Phi_K(x;a)=\tanh(x-a)$. The corresponding metric functions on the two-dimensional moduli space read
\begin{gather*}
g_{\alpha \alpha} = \frac{8}{3}+ \frac{8}{\sinh^2(2\alpha)} \left( \frac{2\alpha}{\tanh(2\alpha)}-1 \right), \\
 g_{\beta \beta} = \frac{8}{3}- \frac{8}{\sinh^2(2\alpha)} \left( \frac{2\alpha}{\tanh(2\alpha)} -1\right),
\end{gather*}
while the effective potential is
\[
V(\alpha)=\frac{8}{3} \frac{17+24\alpha +9(-1+8\alpha){\rm e}^{4\alpha} -9{\rm e}^{8\alpha}+{\rm e}^{12\alpha}}{(-1+{\rm e}^{4\alpha})^3}.
\]
We plot these functions in Figure~\ref{metric-plot}, left panel. It is clearly visible that we encounter the singularity. The moduli space metric near the annihilation point reads
\[
{\rm d}s^2_{\phi^4}=\frac{16}{3}{\rm d}\alpha^2 + \frac{64 \alpha^2}{15}{\rm d}\beta^2
\]
and, therefore, there is a zero of the metric function $g_{\beta\beta}$ at $\alpha=0$, which coincides with the singularity of the reduced effective potential.

Similar results can be obtained for the sine-Gordon model where
\[
U=1-\cos \phi.
\]
Now, the kink is $\Phi_K(x;\alpha)=4\arctan {\rm e}^{x-\alpha}$. The metric functions are
\[
g_{\alpha \alpha}= 16+ \frac{32\alpha}{\sinh(2\alpha)}, \qquad
 g_{\beta \beta} = 16 - \frac{32\alpha}{\sinh(2\alpha)},
\]
while the effective potential is
\[
V(\alpha)=16\left[ 1-\frac{1}{2\cosh^2(\alpha)} \left( 1+\frac{2\alpha}{\sinh(2\alpha)} \right)\right],
\]
see Figure~\ref{metric-plot}, right panel. Once again, the metric function $g_{\beta \beta}$ has a zero at $\alpha=0$, which translates into a divergence of the reduced effective potential.

These examples show that the moduli space with nonzero CoM momentum splits into two separated parts. Solutions are trapped in the branch with positive or negative $\alpha$, and no transition is possible. This means that, e.g., while the breather at rest has a very good description within the CCM approximation in the sine-Gordon model, at least for the non-relativistic regime, its boosted version meets a serious obstacle.

\subsection[AKK collisions in the phi\^{}6 model]{AKK collisions in the $\boldsymbol{\phi^6}$ model}
Now we switch to a theory with non-symmetric kinks where we obtain a wormhole geometry for the kink collisions. The simplest example is provided by the $\phi^6$ model
\[
U=\frac{1}{2} \phi^2\bigl(1-\phi^2\bigr)^2,
\]
where the kink and antikink connect the (0,1) and (1,0) vacua, respectively. They read
\[
\Phi_K(x;\alpha)=\sqrt{\frac{1+\tanh(x-\alpha)}{2}}, \qquad \Phi_{\bar{K}}(x;\alpha)=\sqrt{\frac{1-\tanh(x-\alpha)}{2}}.
\]
Crucially, we cannot reach the true vacuum on the moduli space since
\[
\Phi_{\bar{K}}(x;\alpha) \neq -\Phi_K(x;\alpha)
\]
for any $\alpha$. In this case the interaction part of the metric functions is
\[
\tilde{M}(\alpha)=\frac{1}{8} \int_{-\infty}^\infty \frac{{\rm d}x}{\cosh^2(x-\alpha)\cosh^2(x+\alpha) \sqrt{1-\tanh(x+\alpha)} \sqrt{1+\tanh(x-\alpha)} }
\]
and it is a function which first monotonically grows from zero at $\alpha=\infty$ to a maximum ${\tilde{M}_0\approx 0.2468}$ at $\alpha_0\approx 0.45$ and then decreases again to zero at $\alpha=-\infty$. As a consequence, we find a positive definite metric everywhere without any singularities. The metric component~$g_{\beta \beta}$ decreases from $g_{\beta \beta}(\infty)=\frac{1}{4}$ (twice the kink mass) to its minimal value $g_{\beta \beta}\approx 0.0032$ at the throat at $\alpha\approx \pm 0.45$. Because this value is small, there will be a large potential energy barrier in the reduced effective potential when $P$ is reasonably sized. Hence, although we do not have a singularity in the theory, the CCM is still likely unphysical, at least for larger $P$.

\section{Adding modes}
We have seen that the minimal moduli space capturing collisions of a {\it symmetric kink and antikink} with different velocities suffers from a singularity, which prevents the solitons from meeting or passing through each other. This happens for arbitrarily small values of the center of mass momentum and, therefore, is an unphysical artifact of this minimal CCM.


We will cure this unwanted phenomenon by adding new degrees of freedom to the CCM, enlarging the dimension of the moduli space. Natural candidates for such new collective coordinates are normal modes or Derrick modes. For concreteness, we perform the construction for the $\phi^4$ model. Here, a single kink has a unique normal mode, also called the shape mode, and we use it as the additional collective coordinate. The restricted set of configurations is, as usual, a sum of a kink and an antikink, now each one with its shape mode included
\begin{gather*}
\Phi_{K\bar{K}}(x; \alpha,\beta, A,B)=-1-\tanh(x-\alpha-\beta)-A\frac{\sinh(x-\alpha-\beta)}{\cosh^2(x-\alpha-\beta)}\\
 \hphantom{\Phi_{K\bar{K}}(x; \alpha,\beta, A,B)=}{}+\tanh(x+\alpha-\beta) + B\frac{\sinh(x+\alpha-\beta)}{\cosh^2(x+\alpha-\beta)}.
\end{gather*}
Once again, $\alpha$ parameterizes the distance between the solitons while $\beta$ allows for motion of the center of mass. $A$, $B$ are shape mode amplitudes for each of the solitons. This set results in a~four-dimensional moduli space $\mathcal{M}[\alpha,\beta,A,B]$.

Our first observation initially seems discouraging: this set of restricted configurations have a~singular metric at $\alpha=0$, which may be viewed as a three-dimensional singular subspace spanned by the remaining three coordinates $\beta$, $A$, $B$. Its existence arises from the fact that at this point $\partial_A\Phi_{K\bar{K}}=-\partial_B\Phi_{K\bar{K}}$. Thus, $g_{AA}=g_{BB}=-g_{AB}$, $g_{\alpha A}=-g_{\alpha B}$ and $g_{\beta A}=-g_{\beta B}$, resulting in the vanishing of the metric determinant for any $\beta$, $A$, $B$.

However, parts of this singularity are similar to the apparent singularity previously found in kink-antikink scattering in the $\phi^4$ model~\cite{MORW-2} (with the center of mass at rest). Therefore, this part can be regularized in the standard fashion via a redefinition of the symmetric amplitude coordinate. Concretely, we introduce new coordinates $C$, $D$
\[
A=\frac{1}{2}(C-D), \qquad B=\frac{1}{2} (C+D).
\]
Then,
\begin{gather*}
\Phi_{K\bar{K}}(x; \alpha,\beta, C,D)=-1-\tanh(x-\alpha-\beta)+\tanh(x+\alpha-\beta) \nonumber \\
\hphantom{\Phi_{K\bar{K}}(x; \alpha,\beta, C,D)=}{}- \frac{C}{2} \left(\frac{\sinh(x-\alpha-\beta)}{\cosh^2(x-\alpha-\beta)} -\frac{\sinh(x+\alpha-\beta)}{\cosh^2(x+\alpha-\beta)} \right) \nonumber
\\
\hphantom{\Phi_{K\bar{K}}(x; \alpha,\beta, C,D)=}{} +\frac{D}{2}\left( \frac{\sinh(x-\alpha-\beta)}{\cosh^2(x-\alpha-\beta)} +\frac{\sinh(x+\alpha-\beta)}{\cosh^2(x+\alpha-\beta)} \right). 
\end{gather*}
Now, expanding the configurations at $\alpha=0$ we find that
\begin{gather*}
\Phi_{K\bar{K}} = -1 + D \frac{\sinh(x-\beta)}{\cosh^2(x-\beta)}+ 2\alpha\bigl(1-\tanh^2(x-\beta)\bigr) \\
\hphantom{\Phi_{K\bar{K}} =}{} -\frac{C\alpha}{\cosh(x-\beta)} \bigl(-1+2\tanh^2(x-\beta)\bigr) +o(\alpha).
\end{gather*}
Hence, $\partial_C\Phi_{K\bar{K}} =0$ at $\alpha=0$. The linear dependence on $\alpha$ suggests that this is the usual null vector problem, leading to the vanishing of four metric components $g_{C\alpha}$, $g_{C\beta}$, $g_{CC}$ and $g_{CD}$. This can be cured by the redefinition $C\to C/\tanh(\alpha)$. Thus, an improved restricted set of coordinates reads
\begin{gather}
\Phi_{K\bar{K}}(x; \alpha,\beta, C,D)=-1-\tanh(x-\alpha-\beta)+\tanh(x+\alpha-\beta) \nonumber \\
\hphantom{\Phi_{K\bar{K}}(x; \alpha,\beta, C,D)=}{}- \frac{C}{2\tanh(\alpha)} \left(\frac{\sinh(x-\alpha-\beta)}{\cosh^2(x-\alpha-\beta)} -\frac{\sinh(x+\alpha-\beta)}{\cosh^2(x+\alpha-\beta)} \right) \nonumber
\\
\hphantom{\Phi_{K\bar{K}}(x; \alpha,\beta, C,D)=}{} +\frac{D}{2}\left( \frac{\sinh(x-\alpha-\beta)}{\cosh^2(x-\alpha-\beta)} +\frac{\sinh(x+\alpha-\beta)}{\cosh^2(x+\alpha-\beta)} \right). \label{4d-set}
\end{gather}
Hence,
\begin{gather}
\Phi_{K\bar{K}} = -1 + D \frac{\sinh(x-\beta)}{\cosh^2(x-\beta)}-\frac{C}{\cosh(x-\beta)} \bigl(-1+2\tanh^2(x-\beta)\bigr) \nonumber\\
\hphantom{\Phi_{K\bar{K}} =}{}+ \frac{2\alpha}{\cosh^2(x-\beta)} +o(\alpha), \label{4d-set-0}
\end{gather}
which has the following metric at the annihilation point ($\alpha =0$),
\[
g_{ij} = \left( \begin{matrix}
\hphantom{-}\dfrac{16}{3} & -\dfrac{D\pi}{2} & \hphantom{-}\dfrac{\pi}{2} & 0 \vspace{2mm}\\
-\dfrac{D\pi}{2} & \dfrac{62 C^2}{21} + \dfrac{14D^2}{15} &-\dfrac{14D}{15} & \dfrac{14C}{15} \vspace{2mm}\\
 \hphantom{-}\dfrac{\pi}{2} & -\dfrac{14D}{15} & \hphantom{-}\dfrac{14}{15} & 0 \vspace{2mm}\\
 \hphantom{-}0 & \hphantom{-}\dfrac{14C}{15} & \hphantom{-}0 & \dfrac{2}{3}
\end{matrix}
\right),
\]
where $\bigl(X^i\bigr)=\bigl(X^1,X^2,X^3,X^4\bigr)=(\alpha,\beta, C,D)$.
This is a rank 4 matrix, except at the point $C=0$ where the rank decreases to 3. Indeed, the determinant at $\alpha=0$ is
\[
\det g_{ij}=\frac{16\bigl(896-45\pi^2\bigr)}{2625} C^2.
\]
The moduli space still has a singularity at $\alpha=C=0$. However, now the singularity is a~two-dimensional surface, spanned by $\beta$ and $D$, immersed in the four-dimensional moduli space. Therefore, the moduli space remains connected and there is no obstacle to pass from positive to negative $\alpha$. The kink and antikink can pass through each other, solving the main issue of the minimal, two-dimensional moduli space.

One can understand the appearance of this singularity using the expansion of the restricted set of configurations at $\alpha=0$, equation~(\ref{4d-set-0}). Then, it is easy to verify that $\partial_\beta \Phi_{K\bar{K}} = D \partial_C \Phi_{K\bar{K}}$ at $C=0$. This implies that $g_{\beta i}$ and $g_{C i}$ are not linearly independent, which leads to the vanishing of the metric determinant.

In addition, we numerically computed the determinant for the full range of the collective coordinates. It vanishes {\it only} for $\alpha=C=0$, providing solid numerical evidence that this moduli space does not contain any other singularities.

To complete the analysis of the CCM in the near-singularity regime we present the effective potential close to $\alpha=C=0$
\[
V(\alpha,C,D)=\frac{28}{15}C^2+\frac{4}{3}D^2+\frac{32}{3}\alpha^2 +2\pi C\alpha,
\]
where we also included the leading order terms in $D$. As in the two-dimensional moduli space case, the potential is bounded at the singularity.

As in the case of the two-dimensional moduli space, we can use the conserved momentum to eliminate the $\beta$ coordinate from the CCM. The final effective reduced potential is
\begin{gather*}
V_{\rm red}(X^a,P)=\frac{1}{2}g^{22} P^2+V(X^a) 
\end{gather*}
with $a=1,3,4$. Near the singularity, where $\alpha=0$, the inverse metric function is
\[
g^{22}=\frac{175}{288} \frac{1}{C^2}
\]
and diverges exactly at the subspace where the moduli space has a singularity, $\alpha=C=0$. This proves that finite energy solutions cannot reach the singularity, exactly as a point particle with nonzero angular momentum cannot reach the origin. It also means that this is a true singularity which cannot be removed by any redefinition of the coordinates.

In order to use the CCM based on (\ref{4d-set}) to study kink-antikink collisions, we have to understand the initial conditions. This has been done in~\cite{MORW}. Clearly, the initial state is a free kink (and antikink) moving with a constant velocity, and this should be described within our CCM. For this purpose, we consider the single kink sector
\[
\Phi_K(x;a,A)=\tanh(x-a)+A\frac{\sinh(x-a)}{\cosh^2(x-a)},
\]
where the second term is the shape mode, and $A$ is its amplitude. The resulting CCM reads
\[
L(a,A)=\frac{1}{2}g_{aa}(A) \dot{a}^2 +\frac{1}{2}g_{AA}\dot{A}^2-V(A),
\]
where
\[
g_{aa}(A)=\frac{4}{3}+\frac{\pi}{2}A+\frac{14}{15}A^2, \qquad g_{AA}=\frac{2}{3}
\]
is again a wormhole type metric on the corresponding moduli space, and
\[
V(A)=\frac{4}{3} +A^2+\frac{\pi}{8} A^3+\frac{2}{35}A^4
\]
is an effective potential. One important result is that there exists a stationary solution describing a kink traveling with a constant velocity $v$. Importantly, it requires a constant nonzero amplitude of the shape mode $A_v$
\begin{gather*}
\dot{a}=v_a, \qquad \frac{1}{2}\left( \frac{\pi}{2} +\frac{14}{15}A_{v_a} \right)v^2_a=2A_{v_a}+\frac{3\pi}{8}A_{v_a}^2 +\frac{8}{35}A^3_{v_a} . \label{1-sol}
\end{gather*}
\begin{figure}[t]\centering
\includegraphics[width=0.32\columnwidth]{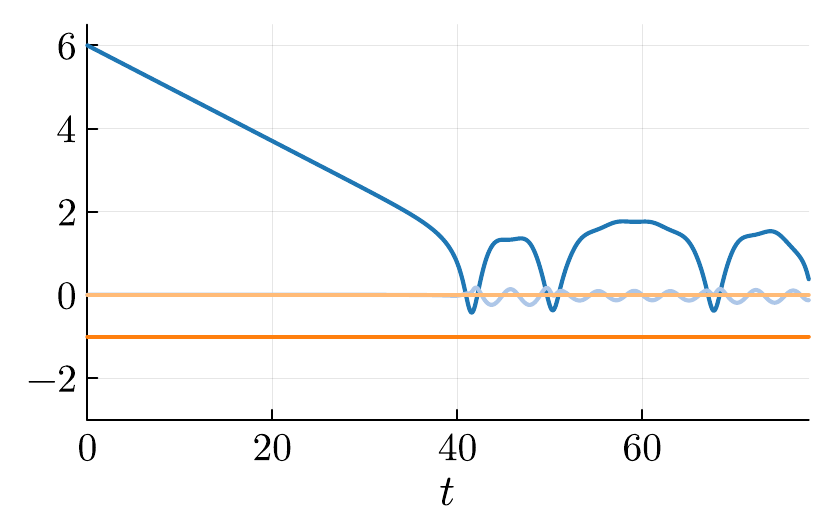}
\includegraphics[width=0.32\columnwidth]{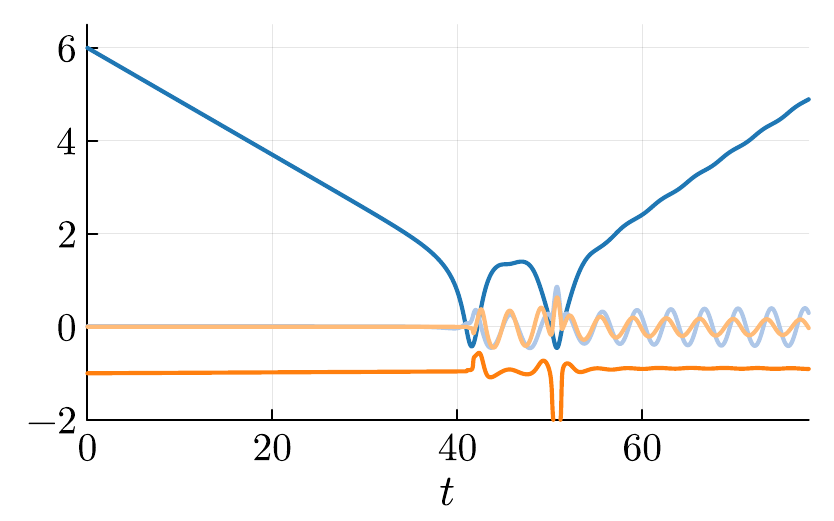}
\includegraphics[width=0.32\columnwidth]{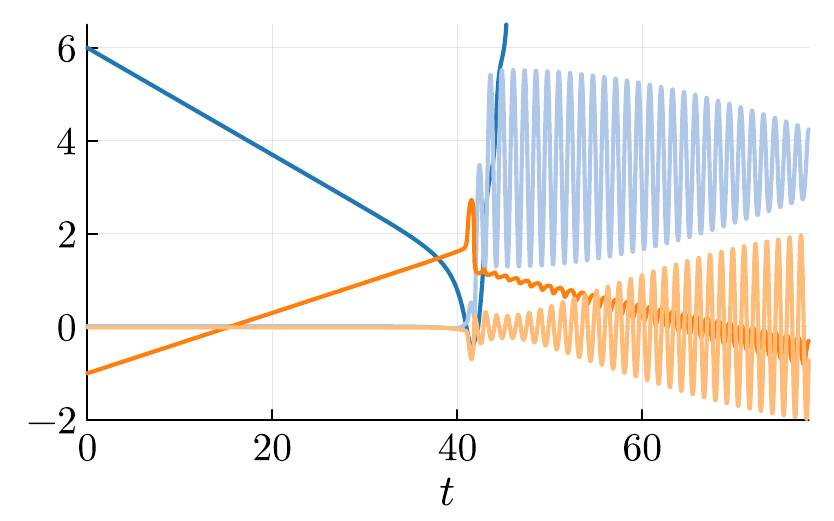}
\includegraphics[width=0.32\columnwidth]{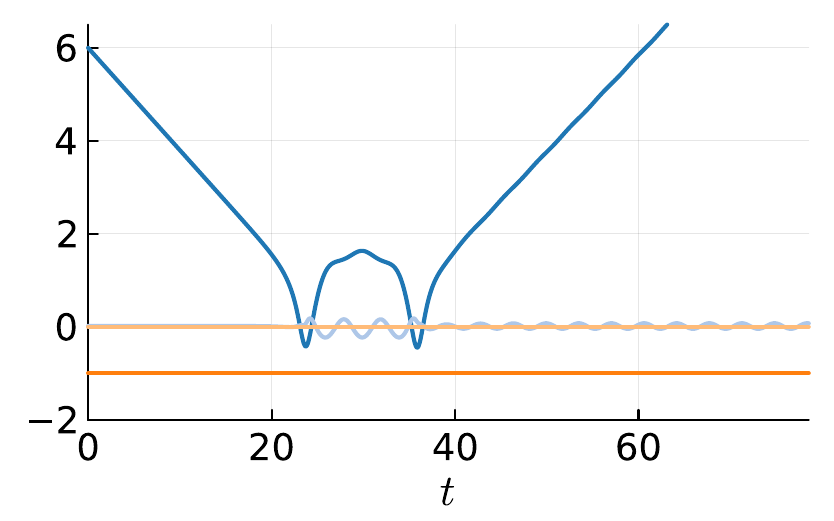}
\includegraphics[width=0.32\columnwidth]{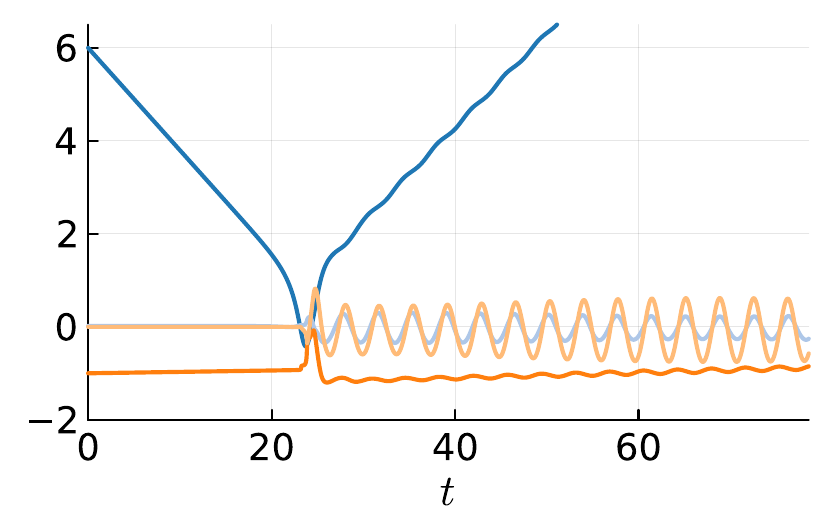}
\includegraphics[width=0.32\columnwidth]{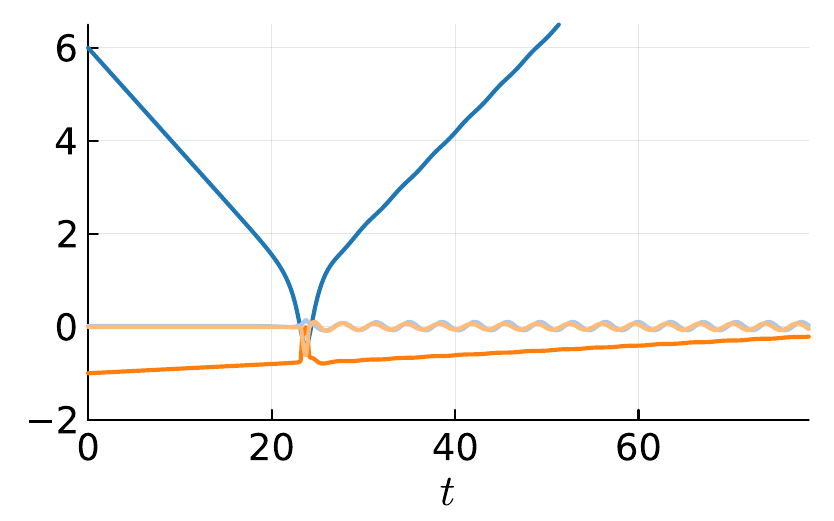}
\includegraphics[width=0.32\columnwidth]{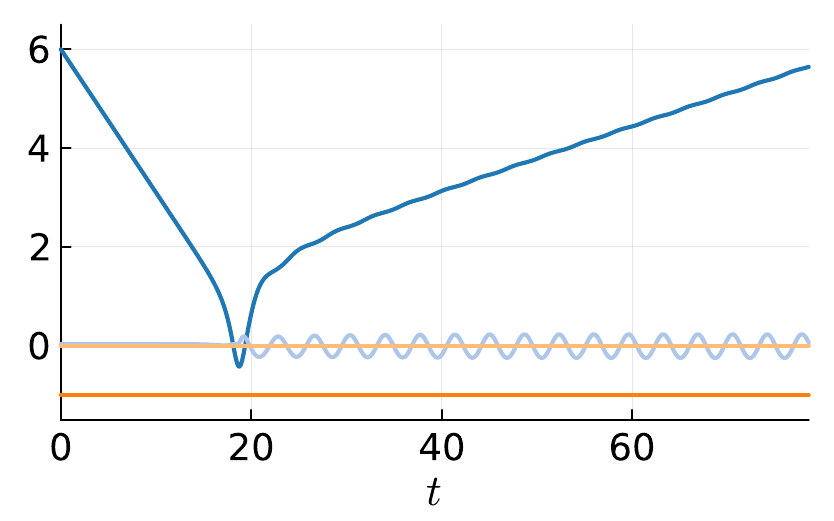}
\includegraphics[width=0.32\columnwidth]{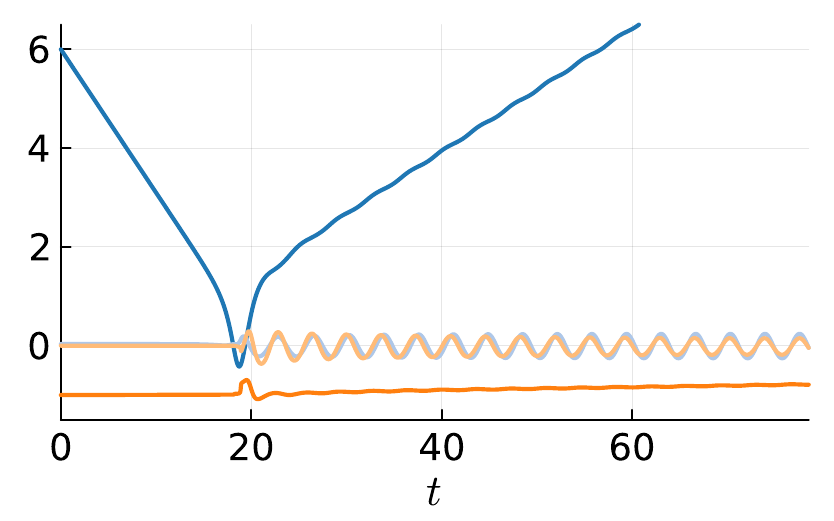}
\includegraphics[width=0.32\columnwidth]{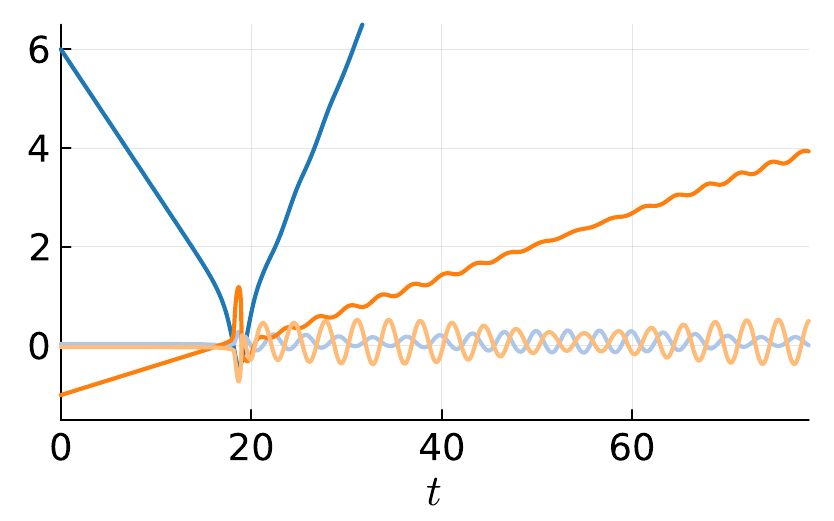}
 \caption{Solutions of the 4-dimensional CCM: $\alpha$ (dark blue), $\beta$ (dark orange), $C$ (light blue), $D$ (light orange). {\it Upper row:} $\dot{\alpha}_{\rm in}=-0.115$. A bion formation case for $\dot{\beta}_{\rm in}=0$ (left) changes into two bounce solution for $\dot{\beta}_{\rm in}=0.001$ (center) and into one bounce solution for $\dot{\beta}_{\rm in}=0.065$ (right). {\it Center row:} $\dot{\alpha}_{\rm in}=-0.22$. A two bounce case for $\dot{\beta}_{\rm in}=0$ (left) changes into one bounce solutions for $\dot{\beta}_{\rm in}=0.003$ (center) and $\dot{\beta}_{\rm in}=0.01$ (right). {\it Lower row:} $\dot{\alpha}_{\rm in}=-0.29$. A one bounce case for $\dot{\beta}_{\rm in}=0$ (left) remains one bounce solutions for $\dot{\beta}_{\rm in}=0.0005$ (center) and $\dot{\beta}_{\rm in}=0.06$ (right).} \label{CCM-run}
 \end{figure}%
Hence, the correct initial conditions are $a(0)=a_0$, $\dot{a}(0)=v_a$, $b(0)=b_0$, $\dot{b}(0)=v_b$ and $A(0)=A_{v_a}$, $B(0)=B_{v_b}$, $\dot{A}(0)=\dot{B}(0)=0$. Here, $a_0$ and $ b_0$ are the initial positions of the antikink and kink, respectively. This easily translates to the initial conditions of the CCM based on the coordinates $\alpha$, $\beta$, $C$, $D$. Namely,
\[
\alpha(0)=\frac{1}{2}(b_0-a_0), \qquad \beta(0)=\frac{1}{2}(b_0+a_0), \qquad \dot{\alpha}(0)=\frac{1}{2}(v_b-v_a), \qquad \dot{\beta}(0)=\frac{1}{2}(v_b+v_a),
\]
and
\[
C(0)=A_{v_a}+B_{v_b}, \qquad D(0)=B_{v_b}-A_{v_a}, \qquad \dot{C}(0)=\dot{D}(0)=0.
\]
As we expected, the CCM does not break down during the evolution, which means that we do not hit the singularity. Furthermore, now the solitons can pass through each other, see, e.g., Figure~\ref{CCM-run} where $\alpha$, which is half of the distance between the kink and antikink, changes its sign.
 \begin{figure}[t]\centering
\includegraphics[width=1.0\columnwidth]{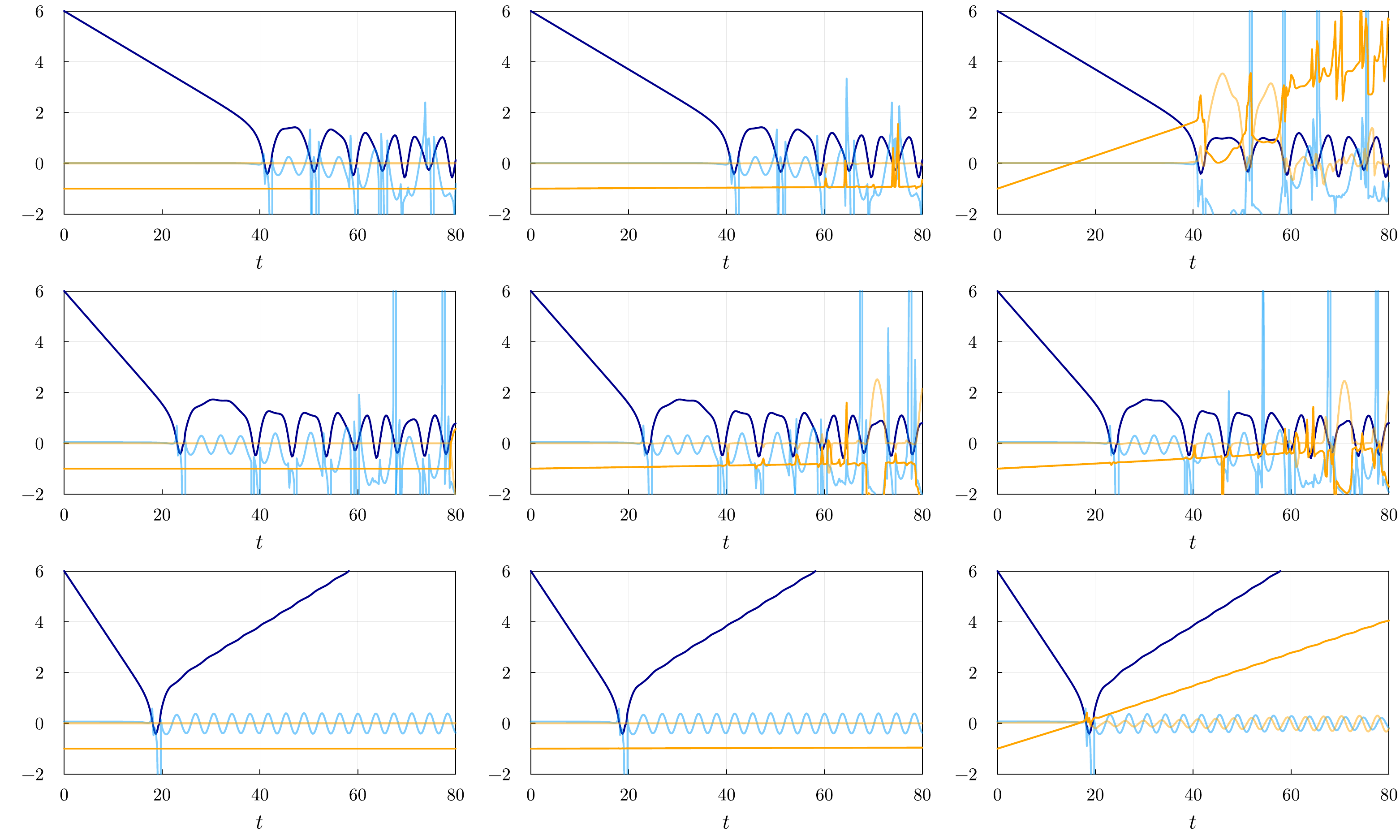}
 \caption{Time dynamics of antikink-kink collisions in the full field theory. $\alpha$ (dark blue), $\beta$ (dark orange), $C$ (light blue), $D$ (light orange). The values of the initial velocities in each panel corresponds with Figure~\ref{CCM-run}.} \label{Full-run}
 \end{figure}%
 In Figure~\ref{CCM-run}, we present some examples of the impact of nonzero momentum on the solutions of the CCM. It is seen that once $P\neq 0$, higher bounce collisions may change into lower order ones or even into annihilation by bion formation. Similarly, a bion formation occurring for $P=0$ can change into one-bounce scattering. In all examples, the solitons temporarily pass through each other during the collision.

 In Figure~\ref{Full-run}, we show the results obtained in the full field theory, where at each time step, the actual field profile is decomposed into the set of configurations (\ref{4d-set}). Definitely, here the impact of the motion of the center of mass is much smaller, which shows that the four-dimensional CCM reveals a much more chaotic character. Thus, although being mathematical sound, this CCM may be less accurate than its counterpart with no motion of the center of mass. Undoubtedly, this issue requires further studies.

\section{Relativistic moduli space with momentum}

Let us now turn to the relativistic moduli space (RMS) framework and include the Derrick scaling parameter as a modulus. However, since we are interested in a non-zero center of mass momentum, we have to modify the usual zero momentum configurations to the following ones
\[
\Phi_{K\bar{K}}(x; a, b, \lambda_a,\lambda_b) = \Phi_K(\lambda_a(x-a))-\Phi_K(\lambda_b(x-b)) +\Phi_v,
\]
where $\lambda_a$, $\lambda_b$ are independent scaling deformations of the solitons.
This can be rewritten using the previously introduced CoM coordinates
\[
\Phi_{K\bar{K}}(x; \alpha, \beta, \lambda, \epsilon) = \Phi_K((\lambda-\epsilon)(x+\alpha -\beta))-\Phi_K((\lambda+\epsilon)(x-\alpha -\beta)) +\Phi_v,
\]
where we introduced $\lambda_{a(b)}=\lambda \mp \epsilon$.

As we previously remarked, the zero-momentum RMS, obtained as a two-dimensional truncation where $\epsilon=0$ and $\beta={\rm const}$, possesses a singularity at $\alpha=0$~\cite{RMS}. In the four-dimensional non-zero momentum case the singularity still exists, however, it is defined by a two-dimensional surface. To see this, we expand the restricted configurations around $\alpha=0$. Thus,
\begin{gather*}
\Phi_{K\bar{K}}(x; \alpha, \beta, \lambda, \epsilon) = \Phi_K((\lambda-\epsilon)(x-\beta))-\Phi_K((\lambda+\epsilon)(x -\beta)) +\Phi_v \nonumber \\
\hphantom{\Phi_{K\bar{K}}(x; \alpha, \beta, \lambda, \epsilon) =}{}+ \alpha \big[ (\lambda-\epsilon) \Phi'_K((\lambda-\epsilon)(x-\beta))+ (\lambda+\epsilon) \Phi'_K((\lambda+\epsilon)(x-\beta) \big].
\end{gather*}
Importantly, the derivative $\partial_\lambda \Phi_{K\bar{K}}$ vanishes identically only if $\alpha=0$ and $\epsilon=0$. Indeed,
\[
\partial_\lambda \Phi_{K\bar{K}} = (x-\beta)\big[ \Phi'_K((\lambda-\epsilon) (x-\beta)) - \Phi'_K((\lambda+\epsilon) (x-\beta)) \big]+o(\alpha).
\]
Furthermore, other derivatives do not vanishes identically and are not linearly dependent. For example,
\[
\partial_\epsilon \Phi_{K\bar{K}} =- (x-\beta)\big[ \Phi'_K((\lambda-\epsilon) (x-\beta)) + \Phi'_K((\lambda+\epsilon) (x-\beta)) \big] +o(\alpha).
\]
This means that the corresponding moduli space has a singularity at $\alpha=\epsilon=0$.

To be more concrete, we will again discuss the $\phi^4$ case. Then, the pertinent restricted set of configurations reads
\[
\Phi_{K\bar{K}}(x; \alpha, \beta, \lambda, \epsilon) = -1 - \tanh((\lambda+\epsilon)(x+\alpha - \beta)) + \tanh((\lambda-\epsilon)(x-\alpha - \beta)).
\]
We expand it at $\alpha=0$ and $\epsilon=0$ which corresponds to a small momentum limit. It gives, up to quadratic terms in $\epsilon$
\begin{gather*}
\Phi_{K\bar{K}}(x; \alpha, \beta, \lambda, \epsilon)  =  -1 + \frac{2\epsilon}{\cosh^2(\lambda(x-\beta))}-  \frac{\alpha}{\cosh^4(\lambda(x-\beta))} \\
\hphantom{\Phi_{K\bar{K}}(x; \alpha, \beta, \lambda, \epsilon)  =}{}\times \bigl( \lambda\bigl(1-4\epsilon^2(x-\beta)^2\bigr) +\lambda \bigl(1+2\epsilon^2(x-\beta)^2\bigr)\cosh 2\lambda (x-\beta) \\
\hphantom{\Phi_{K\bar{K}}(x; \alpha, \beta, \lambda, \epsilon)  =\times \bigl(}{} -  2\epsilon^2 (x-\beta) \sinh 2\lambda (x-\beta) ) \bigr),
\end{gather*}
which leads to the following moduli space metric
\[
g_{ij} = \left( \begin{matrix}
 \dfrac{16}{3} \lambda - \dfrac{16 \pi^2}{45} \dfrac{\epsilon^2}{\lambda} & -\dfrac{8 \epsilon}{3} & 0 & 0 \vspace{2mm}\\
 -\dfrac{8 \epsilon}{3} & \hphantom{-}\dfrac{16 \pi^2 \epsilon^2 }{45 \lambda} & 0 &0 \vspace{2mm}\\
\hphantom{-}0 & \hphantom{-}0 & \left( \dfrac{28\pi^4}{225} - \dfrac{32}{5}\right) \dfrac{\epsilon^2}{\lambda^5} & -\dfrac{2(\pi^2-6)}{3} \dfrac{\epsilon}{\lambda^4}\vspace{2mm} \\
 \hphantom{-}0 & \hphantom{-}0 & -\dfrac{2(\pi^2-6)}{3} \dfrac{\epsilon}{\lambda^4} & \hphantom{-}\dfrac{4(\pi^2-6)}{9} \dfrac{\epsilon}{\lambda^3}
\end{matrix}
\right).
\]
Here $\bigl(X^i\bigr)=\bigl(X^1,X^2,X^3,X^4\bigr)=(\alpha,\beta, \lambda, \epsilon)$ and we put $\alpha=0$. It can be checked easily that the metric is of maximal rank unless $\epsilon=0$.

Of course, the effective potential $V\bigl(X^i\bigr)$ has no singularities for any finite values of the collective coordinates and is zero at the singularity.

Due to the Poincar\'e symmetry, nothing can depend on the position of the CoM $\beta$ and, once again, the corresponding momentum $P$ is conserved. As before, it can be used to eliminate the $\beta$ coordinate from the CCM. This leads to the effective reduced potential $V_{\rm{red}} = g^{22} P^2/2 + V(X^a)$, where $X^a=(
\alpha, \lambda,\epsilon)$ and, again, the inverse metric function diverges at the singularity. Namely, for $\alpha=0$, it behaves as
\[
g^{22}= \frac{657}{4(60\pi^2-225)} \frac{\lambda}{\epsilon^2}.
\]
This shows that for any finite energy solution of the CCM with a nonzero momentum it is impossible to hit the singularity, exactly as for the CCM discussed in the previous section.

These considerations lead to the surprising observation that, when the relativistic moduli space framework is combined with the moving CoM, each ingredient cures the disease of the other one. That is to say,
the introduction of the two Derrick moduli $\lambda_a$, $\lambda_b$ cures the singularity of the moving CoM moduli space discussed in Sections~\ref{sec2} and~\ref{sec3} of the present paper. On the other hand, the introduction of the moving CoM modulus cures the singularity of the (non-perturbative) relativistic moduli space discussed in~\cite{RMS} and briefly mentioned above. A detailed numerical investigation of the resulting four-dimensional CCM and its virtues and limitations would be interesting but goes beyond the scope of the present paper.

\section{Summary}

In the present work, we have analyzed the collective coordinate models (CCM) for kink-antikink collisions where the colliding solitons possess different velocities. This means that the center of mass of the solitons has a nonzero momentum. While in the original Lorentz invariant field theory it is simple to go into the CoM frame, in the CCM framework even an infinitesimal CoM momentum leads to unphysical results. The cure required the introduction of additional collective variables.

The first main result is that, generically, the moduli space of the simplest CCM with two collective coordinates, which parametrize the intersoliton distance $\alpha$ and the position of the center of mass $\beta$, has a geometry of a {\it wormhole} with the positive, $\alpha>0$, and negative, $\alpha<0$ parts joined smoothly by a throat with size $R$. When $R>0$, the CCM trajectories may pass through the annihilation point $\alpha=0$. But for {\it symmetric solitons}, found in theories such as $\phi^4$ and sine-Gordon and whose CCM contains the vacuum, the throat of the wormhole shrinks to a point for nonzero CoM momentum, and a {\it singularity} forms. Now, the moduli space consists of two parts which, close to the singularity, look like a double cone.

Dynamically, this means that solutions of the CCM with nonzero momentum are confined to only one part of the moduli space and can never pass through the annihilation point. This unphysical result occurs for any, even arbitrarily small, nonzero values of the momentum, suggesting that something is badly wrong with the CCM approximation. A similar, but less extreme, problem occurs for CCMs which contain a configuration close to the vacuum, like the $\phi^6$ theory. Here, the wormhole throat is very small, creating a large effective energy barrier to cross the point. The singular metric can be understood in analogy to the motion of a point particle on a plane. Indeed, we have shown that the collective coordinates $\alpha$, $\beta$ play the role of radial coordinates $(r,\theta)$, where the momentum of the center of mass is like the angular momentum of the point particle.

It is quite intriguing that exactly the same type of singularity exists in the relativistic moduli space framework~\cite{RMS}. Indeed, to obtain a CCM allowing for relativistic Lorentz contraction of a moving soliton, one has to consider two collective coordinates, the position $\alpha$ and the scale deformation $\lambda$~\cite{Caputo, Rise}. In the case of collisions of a symmetric kink and antikink, it leads to a~moduli space which reveals a singularity where the solitons annihilate each other, i.e., at $\alpha=0$. To circumvent this issue and apply such a semi-relativistic approach to multi-kink collisions, a~regularization scheme was proposed, where the scaling deformation is treated perturbatively $\lambda=1+\epsilon$. Then, each term in the expansion of kink-antikink configurations in the small parameter $\epsilon$ is treated as an independent internal degree of freedom (Derrick mode) and is multiplied by its own amplitude, i.e., a new collective coordinate. This procedure led to CCMs which apparently model the true multi-kink dynamics quite accurately, see for example~\cite{RMS, MORW}. However, this approach does not seem to be applicable to the current problem, since the position of the CoM cannot be treated as a small parameter. \looseness=1

Nonetheless, there exists a nontrivial resolution to the problem of the singularity, which is the second main result of the present work. It requires the inclusion of new collective coordinates, basically internal modes of the solitons, which enlarge the dimensionality of the moduli space. The singularity still remains, but now it is not a one-dimensional line in a two-dimensional moduli space, but a two-dimensional surface in a four-dimensional moduli space. More precisely, this has been found after a suitable redefinition of the collective coordinates which cures the usual null vector problem arising from the symmetric part of the normal mode. Crucially, for nonzero momentum, this singularity does not divide the moduli space into disconnected parts allowing the solitons to meet and pass through each other. We remark that these additional collective coordinates could be normal modes (the shape mode in $\phi^4$ model) or, if no such modes exist, Derrick modes (e.g., the sine-Gordon model). This may provide a CCM for the moving sine-Gordon breather.

The four-dimensional CCMs obtained in this work are mathematically sound and do not suffer from the issue of the singularity which splits the minimal two-dimensional moduli space into dynamically disconnected parts. Nevertheless, their applicability to the modeling of collisions in the full field theory should be further investigated.

This work adds to a growing literature on collective coordinate models of non-BPS soliton systems. There is one result in common for all of the works: if a single soliton has $n$ degrees of freedom, the CCM which describes the dynamics of $N$ solitons requires \emph{more} than $Nn$ collective coordinates. A kink-antikink naively requires just the positional coordinates, but a shape mode or a Derrick mode is needed to describe the true dynamics~\cite{RMS, MORW}. A baby skyrmion has three zero modes, but at least seven are needed to describe 2-baby scattering~\cite{QSSM}. A skyrmion has six zero modes but at least fourteen are needed to describe 2-skyrmion scattering~\cite{2Sk}. All these works point to the same conclusion: it is not simple to deduce the number of modes required to describe non-BPS soliton dynamics, and surprising things can happen if you pick the wrong number.

\subsection*{Acknowledgements}

The authors acknowledge financial support from the Ministry of Education, Culture, and Sports, Spain (Grant No.~PID2020-119632GB-I00), the Spanish Consolider-Ingenio 2010 Programme CPAN (CSD2007-00042), the Xunta de Galicia (Grant No.~INCITE09.296.035PR and Centro singular de investigaci\'on de Galicia accreditation 2019-2022), and the European Union ERDF. C.H.\ is supported by the Carl Trygger Foundation through the grant CTS 20:25.
K.O., T.R., and A.W.\ were supported by the Polish National Science Centre (Grant No.~NCN 2019/35/B/ST2/00059). AW thanks Jose Queiruga for discussion.

\pdfbookmark[1]{References}{ref}
\LastPageEnding

\end{document}